\documentclass[prl,twocolumn,showpacs,preprintnumbers,amsmath,amssymb]{revtex4}
\usepackage{graphicx}
\usepackage{dcolumn}
\usepackage{bm}
\def\Term#1 #2 #3/{\mbox{$\,^{#1}\!#2_{#3}$ }}
\def\Termo#1 #2 #3/{\mbox{$\,^{#1}\!#2^o_{#3}$ }}

\begin{document}
\title{Spectroscopy of the $^1\!S_0-{}^3\!P_0$ Clock Transition of $^{87}$Sr in an Optical Lattice}

\author{Masao Takamoto}
\author{Hidetoshi Katori}
\altaffiliation[Corresponding author :]{ katori@amo.t.u-tokyo.ac.jp}
\affiliation{%
Engineering Research Institute, The University of Tokyo, and PRESTO, Japan Science and Technology Corporation, Bunkyo-ku, Tokyo 113-8656, Japan}%

\date{\today}
\begin{abstract}
We report on the spectroscopy of the $5s^2 \,{}^1\!S_0~(F=9/2) \rightarrow 5s5p \,{}^3\!P_0~(F=9/2)$ clock transition of ${}^{87}{\rm Sr}$ atoms (natural linewidth of 1 mHz) trapped in a one-dimensional optical lattice.
Recoilless transitions with a linewidth of 0.7 kHz as well as the vibrational structure of the lattice potential were observed. By investigating the wavelength dependence of the carrier linewidth, we determined the magic wavelength, where the light shift in the clock transition vanishes, to be $813.5\pm0.9$ nm.
\end{abstract}
\maketitle

Precision spectroscopy of atoms and molecules has been the essential basis for quantum physics: An increase in precision revealed newer aspects of physics, which motivated the continuous endeavor to develop advanced spectroscopic methods \cite{Luiten}.
Thanks to the state-of-the-art frequency synthesis technology \cite{Comb,Stenger2002} that links optical frequencies at $10^{-18}$ accuracy \cite{Stenger2002}, a stringent comparison of the stability and accuracy among optical-clocks \cite{Comb} becomes feasible, which allows for a more accurate definition of the SI second and determination of the fundamental constants \cite{hydrogen}, and leads to the study of time-variation of fundamental constants \cite{fluct,fluct1, fluct2}. 
The existing microwave or optical clocks based on cesium atoms in a fountain \cite{Comb,Cs}, a single mercury ion in a Paul trap \cite{Hg} or ultracold neutral calcium atoms in free fall \cite{Ca}, have so far demonstrated a fractional accuracy and stability on the order of $10^{-15}$ \cite{Comb,Cs,Hg,Ca,Luiten}, and vigorous efforts are being made for their further improvement.

In quest of the novel scheme for future optical standard, we have explored the feasibility of an ``optical lattice clock" \cite{Katori2002,Ido2003,Katori2003}, in which millions of neutral atoms trapped in an engineered optical lattice serve as quantum references effectively free from light field perturbations \cite{Katori1999}.
In this scheme, sub-wavelength confinement of atoms provided by an optical-lattice with less than unity occupation \cite{lattice} completely eliminates both the Doppler and collisional shifts, which are known to cause major uncertainties in optical standards with freely falling atoms; recent study assuming ultracold Ca atoms predicted their contribution could approach down to $8\times 10^{-16}$ in the future \cite{Ca}.
Furthermore, an extended interrogation time of atoms trapped in the lattice potential will provide a $2-3$ orders of magnitude higher line $Q$-factor than that has been obtained for freely falling atoms \cite{Comb,Ca,Ertmer1998}.
With the line $Q$-factor competitive to a single trapped ion clock \cite{Hg}, but with millions of atoms interrogated simultaneously as in the case of neutral-atom-based optical clocks \cite{Comb,Ca,Ertmer1998}, the ``optical lattice clock" is expected to exhibit an exceptionally high stability: Its anticipated accuracy of $10^{-17}$ \cite{Katori2003} can be reached within only 1 second of interrogation time.

In this Letter, we report on the first demonstration of the ``optical lattice clock" by performing spectroscopy on the $5s^2~{}^1\!S_0\ (F=9/2)-5s5p~{}^3\!P_0\ (F=9/2)$ clock transition of $^{87}$Sr atoms trapped in a 1D optical lattice.
In our primary search for the 1-mHz-narrow clock transition, a quenching technique \cite{Diedrich1989} was applied to the $^3\!P_0$ metastable state to artificially broaden the clock transition linewidth. 
We then discuss the method to adjust the light shifts in the clock transition by monitoring the lineshape of the carrier spectrum.

The transition frequency $\nu$ of atoms exposed to a dipole trapping \cite{Ashkin} laser with a slowly varying electric field amplitude of $\mathcal{E}$ is described as \cite{Katori2003}
\begin{equation}\label{l1}
 \hbar \nu = \hbar \nu^{(0)}-\frac{1}{4}\Delta \alpha ({\bf
    e},\lambda )\mathcal{E}^2-\frac{1}{64}\Delta \gamma ({\bf
    e},\lambda)\mathcal{E}^4-\dots,
\end{equation}
where $\nu ^{(0)}$ is the transition frequency between the unperturbed atomic states, $\Delta \alpha ({\bf e},\lambda)$ and $\Delta \gamma({\bf e},\lambda)$ are the differences between the ac polarizabilities and hyperpolarizabilities of the upper and lower states, which depend both on the light polarization unit vector ${\bf e}$ and on the trapping laser wavelength $\lambda$, and $\hbar$ is the Planck constant.
In order to reduce the polarization dependent light shift \cite{Ido2003} and other systematic shifts, we proposed to use the ${}^1\!S_0- {}^3\!P_0$ transition of $^{87}$Sr \cite{Katori2002} that is free from electronic angular momentum.
At the ``magic wavelength" that sets the differential dipole polarizability to zero, the residual contributions due to the polarization dependent light shift, the hyperpolarizability, and the higher-order multipole corrections to the polarizability, were calculated to be less than $10^{-17}$ \cite{Katori2003} for a light field intensity of $I_L={\rm 10\,kW/cm^2}$.
In the lattice scheme, this residual light shift poses the major limitation for the attainable accuracy: The suppression of the Doppler and collisional shifts by the lattice potential, therefore, will bring about more than an order of magnitude better accuracy for neutral atom clock.

\begin{figure}
\includegraphics[width=0.9\linewidth]{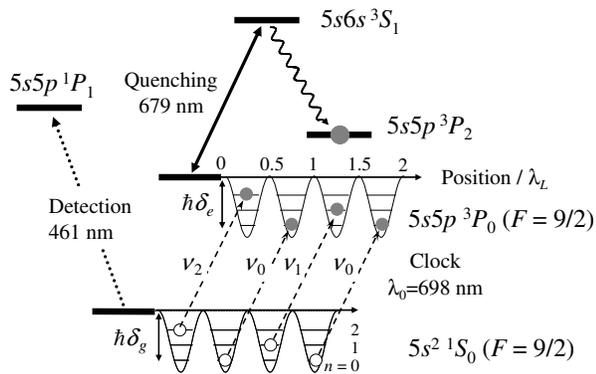}
\caption{Energy levels for the spectroscopy. The $5s^2~{}^1\!S_0$ and $5s5p~{}^3\!P_0$ state are coupled to the upper respective spin states by an off-resonant standing wave light field to produce equal amount of light shifts ($\hbar \delta_g$ and $\hbar \delta_e$) in the clock transition and to provide atoms with sub-wavelength confinement. 
The excited atoms on the $|^1\!S_0\rangle \otimes |n\rangle\rightarrow|^3\!P_0\rangle \otimes |n \rangle$ electronic-vibrational transitions in the light shift potentials were quenched into the $^3\!P_2$ metastable state via the rapidly decaying $^3\!S_1$ state.
The clock transition was sensitively monitored on the ${}^1\!S_0-{}^1\!P_1$ transition.
}
\label{level}
\end{figure}

The relevant energy levels are shown in Fig.~\ref{level}. The $5s^2~{}^1\!S_0\ (F=9/2)-5s5p~{}^3\!P_0\ (F=9/2)$ transition with a hyperfine induced decay rate of $\gamma_0=2\pi \times 1\,{\rm mHz}$ \cite{Katori2002,Kluge} at $\lambda_0=698\ {\rm nm}$ is used as clock transition.
By optically coupling the $5s^2\ ^1\!S_0$ and $5s5p\ ^3\!P_0$ state to the upper respective spin states with a standing-wave trapping laser tuned to the ``magic wavelength", lattice potentials \cite{lattice} with equal depth are produced for the electronic states of the clock transition \cite{Katori2003}, which tightly confine atoms to the sub-wavelength region, the so-called Lamb-Dicke regime \cite{Dicke}. This configuration enables recoil-free as well as Doppler-free spectroscopy, as has been demonstrated on the  ${}^1\!S_0-{}^3\!P_1$ transition of $^{88}$Sr atoms \cite{Ido2003}.

\begin{figure}
\includegraphics[width=0.9\linewidth]{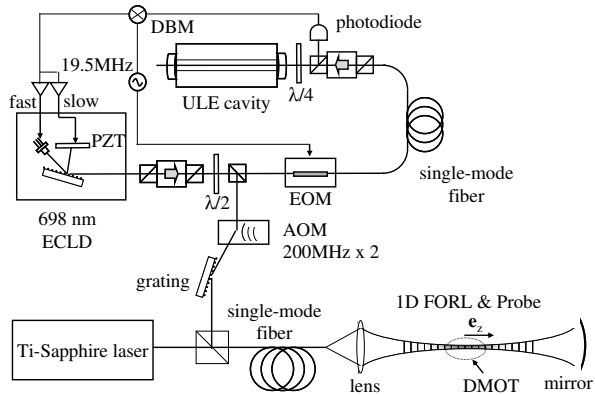}
\caption{ An external cavity laser diode (ECLD) was electronically stabilized to a high-finesse reference cavity made of ultra-low-expansion (ULE) glass. This clock laser at $\lambda_0=698\,{\rm nm}$ was superimposed on the Ti-Sapphire laser (used for trapping atoms) and coupled into a polarization-maintaining single-mode fiber. The radiation from the fiber was focused onto the ultracold atom cloud and the trapping beam alone was retro-reflected to form an optical lattice, where ${\bf e}_z$ denotes the unit vector parallel to the clock and the trap laser axes.
}
\label{setup}
\end{figure}

Figure \ref{setup} depicts the experimental setup.
In order to probe the clock transition, an external cavity laser diode (ECLD) was electronically stabilized to a high-finesse reference cavity made of ultra low expansion (ULE) glass using an FM sideband technique \cite{Hall1983,Bergquist1999}. 
The output of the stabilized laser was frequency-tuned by an acousto-optic modulator (AOM) and spectrally filtered by a grating to remove the amplified spontaneous emission (ASE) from the ECLD, as it may excite Sr transitions in the spectral range of $679-707\,{\rm nm}$ that share the same electronic states used for the clock transition.
This clock laser was then superimposed onto the trapping laser (Ti-Sapphire laser, Coherent 899) and coupled into a polarization-maintaining single-mode fiber.

Ultracold $^{87}{\rm Sr}$ atoms were produced  as described previously \cite{Mukaiyama2003}.
The atoms were precooled on the $^1\!S_0-{}^1\!P_1$ transition at $\lambda=461\ {\rm nm}$. During the precooling process, the atoms that leaked into the $5s5p\ ^3\!P_2$ metastable state via the $5s4d\ ^1\!D_2$ state were recycled to the $^1\!S_0$ ground state via the $5s5p\,^3\!P_1$ state by exciting the $5s5p\ ^3\!P_2 \rightarrow 5s6s\ ^3\!S_1$ and the $5s5p\ ^3\!P_0 \rightarrow 5s6s\ ^3\!S_1$ transition \cite{Mukaiyama2003}.
In about 300 ms the number of the precooled atoms reached 90~\% of its steady state population.
After turning off the precooling and the pumping lasers, we started the narrowline cooling on the $^1\!S_0-{}^3\!P_1$ transition by employing the DMOT scheme \cite{Mukaiyama2003}.
We first frequency-modulated trapping and molasses lasers for 70 ms to increase the velocity capture range, then turned off the modulation and reduced the laser intensity to optimize the atom loading into an optical lattice.
The optical lattice formed by the standing wave of linearly polarized light at $\lambda_L\sim800~{\rm nm}$ was kept on during the experimental sequence. 
The lattice laser with the 1/e beam waist of $w_0=16\ \mu{\rm m}$ was focused onto the ultracold atom cloud of the narrowline MOT to load roughly $10^5$ atoms into this far-off-resonant lattice (FORL) in 70 ms.
This lattice potential typically provides a harmonic oscillation frequency of $\Omega/2 \pi \approx 70\ {\rm kHz}$ in the axial direction ${\bf e}_z$ (see Fig.~\ref{setup}) as measured from the first sideband spectrum, corresponding to the effective peak laser intensity of $I_L\approx 30\,{\rm kW/cm}^2$.

In search for the 1 mHz narrow $^1\!S_0-{}^3\!P_0$ transition \cite{Lemonde2003}, we saturation-broadened the clock transition: By guiding the clock  and trap laser in the same optical fiber as shown in Fig.~\ref{setup}, we tightly focused the clock  laser exactly at the trapped atom cloud in the lattice. We thus achieved a peak power density of $I_p=120\ {\rm W/cm}^2$ by focusing 1 mW of laser power onto a trapping beam radius of approx. $w_0$. With this intensity a saturation broadening of $\gamma_0\sqrt{1+I_p/I_0}\approx 2 \pi \times 18~{\rm kHz}$ was obtained, where $I_0=0.4\ {\rm pW/cm}^2$ is the saturation intensity of the hyperfine induced $^1\!S_0~(F=9/2)-{}^3\!P_0~(F=9/2)$ transition.
By guiding both lasers in the same fiber, the wavevector of the clock laser with $|{\bf k}_0|=2\pi/\lambda_0$ was intrinsically aligned parallel to the fast axis ($\parallel {\bf e}_z$) of the lattice potential, which gave the Lamb-Dicke parameter for this axis $\eta=|{\bf k}_0| \sqrt{\hbar/2 m \Omega}\approx 0.26$, with $m$ the mass of the $^{87}{\rm Sr}$ atom.
We facilitated the first search for the narrow transition by frequency modulating the clock laser with a spectrum spread of tens of MHz. 
In addition, by optically coupling the $^3\!P_0$ state to the rapidly decaying ${}^3\!S_1$ state with $\lambda=679\ {\rm nm}$ laser radiation, we quenched the $^3\!P_0$ state lifetime and simultaneously transferred the population into the long-lived $^3\!P_2$ metastable state \cite{Derevianko2001,Yasuda2003} via the ${}^3\!S_1$ state where the branching ratio for the decay to the $^3\!P_2$ state is 5/9.
We typically irradiated both the clock and quenching lasers for 100 ms on trapped atoms. We then excited the dipole allowed ${}^1\!S_0-{}^1\!P_1$ transition at 461 nm with a linewidth of $\gamma=2\pi\times 32\ {\rm MHz}$ (Fig.~\ref{level}) for 1 ms to observe fluorescence photons proportional to the number of atoms remaining unexcited in the $^1\!S_0$ ground state.
This shelving technique \cite{shelving} allowed the observation of the clock transition with unit quantum efficiency. We note that Courillot {\it et al.} \cite{Lemonde2003} recently reported the $^1\!S_0-{}^3\!P_0$ transition frequency  by observing 1.4-MHz-wide Doppler profile with 1 \% excitation, in which they first looked for the clock transition by cascading three allowed transitions.

\begin{figure}
\includegraphics[width=0.8\linewidth]{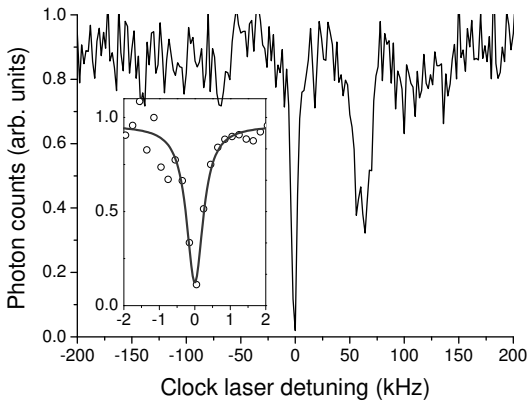}
\caption{The ground state population as a function of the clock laser detuning measured at lattice laser wavelength $\lambda_L=815\,{\rm nm}$.
The upper and the lower sidebands at $\approx \pm 64\ {\rm kHz}$ correspond to the heating and the cooling sidebands, respectively.
The base line fluctuation of $\approx 15 \%$ is due to the shot-to-shot fluctuation of atoms loaded into the optical lattice.
The inset shows the recoilless spectrum (the carrier component) with a linewidth of 0.7 kHz (FWHM) measured at $\lambda_L=813$ nm.
}
\label{spectrum}
\end{figure}

After the first detection of the clock transition with a linewidth of tens of MHz, we gradually decreased the intensities of the clock and the quenching lasers to observe a narrower resonance line.
In order to completely remove the light shift and broadening due to the quenching laser, we alternately chopped the clock and quenching lasers: After irradiating the clock laser for 5 ms, we applied the quenching laser. We repeated this sequence several times to increase the depletion of the ${}^1\!S_0$ state by exciting the clock transition.
Figure \ref{spectrum} shows the $^1\!S_0$ state population as a function of the clock  laser detuning.
Each data point was measured in a single cycle with roughly $10^5$ atoms in the lattice, which required half a second in total for cooling, capturing, and detecting atoms. 
Thanks to the quenching laser that transferred the atom population into the $^3\!P_2$ metastable state, nearly 100 \% excitation of the ground state population was observed for the carrier component.
The upper and the lower sideband at $\pm \Omega/2\pi\approx \pm 64\ {\rm kHz}$ corresponds to the  $|^1\!S_0\rangle \otimes |n\rangle\rightarrow|^3\!P_0\rangle \otimes |n\pm1\rangle$ transitions, where $\Omega$ and $|n\rangle$ denotes the oscillation frequency and the vibrational state of atoms in the lattice potential.
The asymmetry in the heating and cooling sidebands \cite{Diedrich1989} inferred a mean vibrational state occupation of $\langle n \rangle\approx 0.5$ or an atom temperature of $T=2.8$ $\mu$K.
The narrowest linewidth of 0.7 kHz, as shown in the inset of Fig.~\ref{spectrum}, was observed at $\lambda_L=813\ {\rm nm}$ by reducing the clock laser intensity down to $I_p=0.1\ {\rm W/cm^2}$, where the saturation broadening of 0.5 kHz was comparable to a clock laser frequency jitter of $\approx0.5$ kHz. 

In this clock transition, the first-order Zeeman shift of $106 \times m_F$ Hz/G for the $\Delta m_F=0$ transition appears due to the hyperfine mixing in the $^3\!P_0$ state \cite{Kluge,Peik1994}. 
In addition, a polarization dependent light shift, which is approximately $ 1\ {\rm Hz/(kW/cm^2)}$ for the $m_F=\pm 9/2$ states and becomes smaller as $|m_F|$ decreases, is present \cite{Katori2003}.
Assuming a stray magnetic field of $B<10$ mG and a lattice laser intensity of $I_L\approx 30\ {\rm kW/cm^2}$, both contributions are less than 40 Hz.
At the moment, therefore, the carrier linewidth is mainly determined by a frequency jitter of the clock laser and a saturation broadening. 
Since we are employing a 1D optical lattice that contained about tens of atoms per a lattice site, collisional frequency shifts might be present, which would ultimately be eliminated by using a 3D optical lattice \cite{lattice} with less than unity occupation \cite{Katori2002,Katori2003}.

In order to determine the ``magic wavelength" for the lattice laser, we measured the wavelength ($\lambda_L$) dependent carrier lineshape, since the mismatch of the confining potentials introduces an additional carrier linewidth broadening as discussed in the following. 
The energy shift of atoms in the $n$-th vibrational state of the $i=e$ (excited) or $g$ (ground) electronic state of the clock transition is written as
\begin{equation}
\hbar \delta_i(n,\lambda_L)=u_i \left( \lambda_L \right)+(n+\frac{1}{2})\hbar \Omega_i (\lambda_L), 
\label{stark}
\end{equation}
where $u_i(\lambda_L) <0$ is the light shift at the anti-node of the standing wave and $\Omega_i(\lambda_L)/ 2\pi \approx \sqrt{-2u_i/m}/\lambda_L $ is the vibrational frequency of atoms in the fast axis of the lattice.
Taking $\nu_n$ the transition frequency for the $|^1\!S_0\rangle \otimes |n\rangle\rightarrow|^3\!P_0\rangle \otimes |n \rangle$ vibrational transition (see Fig.~\ref{level}), Eq.~(\ref{stark}) infers that the transition frequency difference  between adjacent vibrational transitions is given by the vibrational frequency difference $\delta \Omega \equiv \Omega_e-\Omega_g$ of the lattice potentials in the excited and the ground state, i.e., $ \nu_{n+1}-\nu_n= \delta \Omega
$.
At a finite temperature $T$, the occupation probability $p_n$ of the atoms in the $n$-th vibrational state obeys the Boltzmann distribution law, i.e., $p_{n+1}/p_n=\exp(-\hbar\Omega_g/k_B T)\equiv f_B$.
Therefore, the carrier spectrum for non-degenerate light shift trap ($\Omega_e\neq\Omega_g$) consists of several Lorentzian excitation profiles with frequency offset given by $\delta \Omega$ and their peak height weighted by the Boltzmann factor $f_B$.

\begin{figure}
\includegraphics[width=0.8\linewidth]{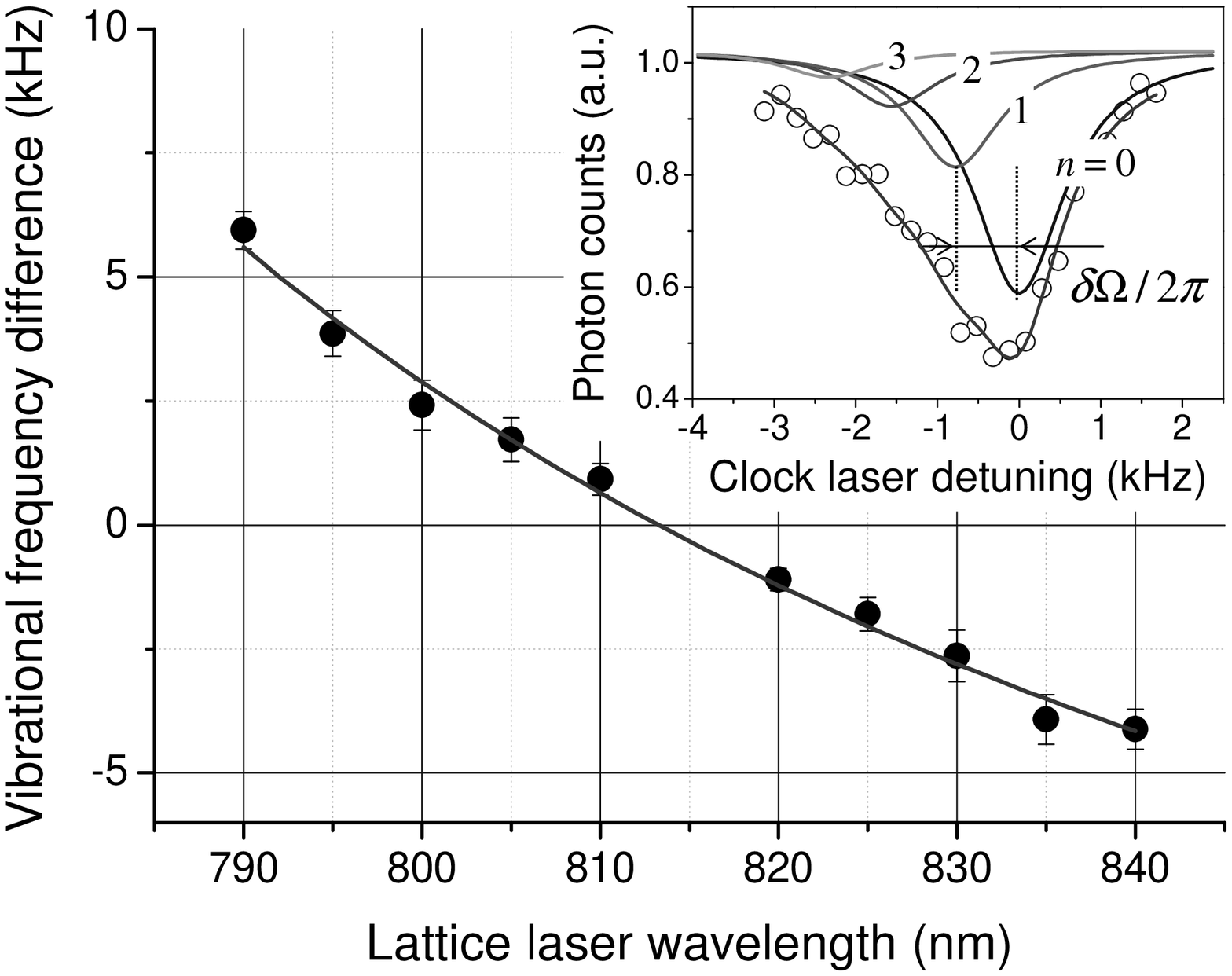}
\caption{The inset shows the carrier spectrum (open circles) for the $|^1\!S_0\rangle \otimes |n\rangle\rightarrow|^3\!P_0\rangle \otimes |n \rangle$ transition measured at lattice laser wavelength $\lambda_L=820$ nm. This lineshape is fitted by 4 Lorentian corresponding to $n=0,1,2,3$ vibrational transitions to deduce the Boltzmann factor $f_B\approx 0.5$ and the differential vibrational frequency $\delta \Omega/2 \pi \approx 0.8$ kHz.
The latter is plotted as a function of lattice wavelength to determine the degenerate wavelength ($\delta \Omega=0$) to be $\lambda_L=813.5\pm0.9$ nm.}
\label{linewidth}
\end{figure}

The inset of Fig.~\ref{linewidth} demonstrates the linewidth broadening of the carrier spectrum at the lattice laser wavelength $\lambda_L=820$ nm: The profile was fitted by 4 Lorentzians corresponding to $n=0,1,2,3$ vibrational states to determine the differential vibrational frequency $\delta \Omega/2\pi=0.8$ kHz and the Boltzmann factor $f_B\approx 0.5$.
By applying this fitting procedure for carrier lineshapes measured at different lattice laser wavelength, we deduced the vibrational frequency mismatch $\delta \Omega/2 \pi$ as shown by filled circles in Fig.~\ref{linewidth}. 
These data points were then interpolated by quadratic polynomial, which approximated the wavelength dependence of the frequency mismatch to find the degenerate wavelength to be $\lambda_L=813.5\pm0.9$ nm.
This is also confirmed by measuring the linewidth reduction of the clock transition around the wavelength.
This degenerate wavelength agreed to the previous theoretical calculation \cite{Katori2003} within 2 \%, in which the discrepancy may be attributed to the truncation in summing up the light shift contributions and to the limited accuracy of the available transition strengths.

In summary, we have demonstrated for the first time the Doppler-free spectroscopy on the $^1\!S_0-{}^3\!P_0$ transition of $^{87}$Sr atoms trapped in a 1D Stark-free optical lattice and determined the magic wavelength.
This demonstration is an important step for the realization of the ``optical lattice clock" \cite{Katori2002,Katori2003} that would provide a significant improvement in the stability over existing optical clocks.

The authors would like to thank K. Okamura and M. Yasuda for their experimental assistance and T. Eichler for careful reading of the manuscript.
H. K. acknowledges financial support from Japan Society for the Promotion of Science under Grant-in-Aid for Young Scientists (A) KAKENHI 14702013.

\end{document}